\documentclass[aps,prd,amsmath,floats,floatfix, twocolumn,superscriptaddress,nofootinbib,showpacs,longbibliography]{revtex4-1}
\usepackage{aas_macros}
\usepackage[T1]{fontenc}
\usepackage[utf8]{inputenc}
\usepackage{lmodern}
\usepackage{verbatim}
\usepackage[dvipsnames, usenames]{xcolor}
\definecolor{linkcolor}{rgb}{0.0,0.3,0.5}
\usepackage[hypertexnames=false, unicode, colorlinks=true, linkcolor=linkcolor,citecolor=linkcolor, filecolor=linkcolor,urlcolor=linkcolor,pdfusetitle]{hyperref}
\usepackage{amsmath,amssymb,graphicx,gensymb}
\usepackage{subfigure}
\usepackage[normalem]{ulem}
\usepackage{mathtools}

\definecolor{orange}{rgb}{1,0.5,0}
\definecolor{mygray}{gray}{0.9}
\definecolor{darkgreen}{rgb}{0.33, 0.42, 0.18}

\allowdisplaybreaks

\begin{document}
\title{A rapid method for preliminary identification of subthreshold strongly lensed counterparts to superthreshold gravitational-wave events}

\author{Srashti Goyal}
\affiliation{International Centre for Theoretical Science, Tata Institute of Fundamental Research, Bangalore - 560089, India}

\author{Shasvath J. Kapadia}
\affiliation{Inter-University Centre for Astronomy and Astrophysics, Post Bag 4, Ganeshkhind, Pune - 411007, India}

\author{Jean-Ren\'e Cudell}
\affiliation{STAR Institute, Universit\'e de Li\`ege, 4000 Li\`ege, Belgium}

\author{Alvin K. Y. Li}
\affiliation{LIGO Laboratory, California Institute of Technology, Pasadena, CA 91125, USA}

\author{Juno C. L. Chan}
\affiliation{Niels Bohr International Academy, Niels Bohr Institute, Blegdamsvej 17, 2100 Copenhagen, Denmark}

\begin{abstract}
Gravitational waves (GWs) from stellar-mass compact binary coalescences (CBCs) are expected to be strongly lensed when encountering large agglomerations of matter, such as galaxies or clusters. Searches for strongly lensed GWs have been conducted using data from the first three observing runs of the LIGO-Virgo GW detector network. Although no confirmed detections have been reported, interesting candidate lensed pairs have been identified. In this work, we delineate a preliminary analysis that rapidly identifies pairs to be further analyzed by more sophisticated Bayesian parameter estimation (PE) methods. The analysis relies on the Gaussian/Fisher approximation to the likelihood and compares the corresponding approximate posteriors on the chirp masses of the candidate pair. It additionally cross-correlates the rapidly produced localization sky areas (constructed by Bayestar sky-localization software). The analysis was used to identify pairs involving counterparts from
targeted sub-threshold searches
to confidently detected super-threshold CBC events. The most significant candidate ``super-sub'' pair deemed by this analysis was subsequently found, by more sophisticated and detailed joint-PE analyses, to be among the more significant candidate pairs, but not sufficiently significant to suggest the observation of a lensed event \cite{o3btechdoc}.
\end{abstract}

\maketitle

\section{Introduction}\label{sec:introduction}

The LIGO-Virgo-KAGRA\cite{advligo, advvirgo,KAGRADet} network of ground-based, interferometric, gravitational-wave (GW) detectors has completed three observing runs O1, O2, and O3. These runs have provided $\sim 100$ detections of compact binary coalescence (CBC) events \cite{gwtc-3, ias-1, ias-2, ias-3, ias-4, ogc-3}. The majority of these correspond to binary black hole (BBH) mergers, although binary neutron star (BNS) \cite{GW170817-DETECTION, GW190425} and neutron star black hole (NSBH) \cite{NSBH-DETECTION} binary mergers have also been observed. 

These detections have enabled many novel probes of various aspects of astrophysics, cosmology, and fundamental physics, including constraints on the populations of stellar-mass compact binaries that merge within Hubble time \cite{gwtc3-population}, distance-ladder-independent measurements of the Hubble constant \cite{gw170817-hubble}, unique tests of general relativity in the strong field regime \cite{gwtc3-tgr}, and the nature of matter at extreme densities via constraints on the neutron star equation of state \cite{gw170817-eos}. Nevertheless, several discoveries involving GWs still remain to be made. Among them is the anticipated observation of gravitationally lensed GWs.
Propagating GWs, like light, will have their paths deviated if they encounter matter inhomogeneities \cite{lensing_gw_1, lensing_gw_2, lensing_gw_3, lensing_gw_4}. In particular, when GWs from stellar mass CBCs, detectable by LIGO-Virgo, encounter galaxies or clusters, they will be strongly lensed, resulting in the possible production of multiple temporally resolvable images \cite{Ng:2017yiu, Dai:2016igl, smith2018, Li:2018prc}. These images are (de)-magnified copies of the source separated by time delays that span minutes to weeks. Thus, they will have identical phase evolution, although their amplitudes will generally differ by a constant factor (see, e.g., \cite{haris2018}). In addition, a constant phase difference of either $0, \pi/2$ or $\pi$, called the Morse phase, will be incurred between the images, depending on the image type (type I, II or III \cite{Dai:2016igl, Ezquiaga2020}).

Several techniques have been devised to search for such strongly lensed pairs of GW events, most of which rely on the identical phase evolution of the images, the superposed GW localization sky areas of the images~\cite{Wong:2021lxf}, and the Morse phase. These include two low-latency techniques. One is a machine learning (ML) based method that compares time-frequency maps and localization sky areas of individual events in a candidate lensed pair \cite{goyal2021}. The other is a posterior-overlap (PO) method that compares the existing Bayesian posterior distribution of the (intrinsic and sky location) parameters of individual events acquired from large-scale Bayesian parameter estimation (PE) exercises \cite{haris2018}. Other more comprehensive though computationally expensive methods involve sampling a joint-likelihood, constructed from the GW-likelihoods of the individual events in the candidate lensed pair \cite{liu2021,golum,hanabi,McIsaac:2019use}. Each of these methods has been employed to search for lensed pairs involving superthreshold/confidently detected GW events from O1, O2, and O3 \cite{gwtc-1,gwtc-2,gwtc-3}. No candidate pair was deemed sufficiently significant to claim detection of GW lensing \cite{LIGOScientific:2021izm,hannuksela2019,theligoscientificcollaboration2023search}.

It has been suggested in the literature that the rate of lensed events -- where one event in a lensed pair is a superthreshold GW event, while the other is a subthreshold event whose reduced significance could be due to one (or more) of several reasons, including de-magnification -- is larger by a factor of few than the rate of lensed pairs where both GW events are superthreshold \cite{wierda2021}. It is therefore worthwhile to search for such ``super-sub'' lensed pairs. A lensed GW counterpart can be missed by the usual matched-filter searches due to its low signal-to-noise (SNR) as a result of lensing de-magnification or reduced sensitivity of the detector for certain parameters of the source. Hence, a separate matched-filter search is done to find the possible subthreshold lensed counterparts for each of the targeted super threshold events \cite{li2019targeted, McIsaac_pycbc}.  This is done by utilizing the posteriors of intrinsic parameters of the superthreshold events, to construct a reduced template bank that enables a deeper search for subthreshold events by reducing the background noise. The candidate super-sub pairs have been found through these searches during O2 \cite{McIsaac_pycbc, Dai2020} and O3 \cite{theligoscientificcollaboration2023search}. However till now, no confident detection has been made.

While all the superthreshold candidates (events having search false alarm rate ($\mathrm{FAR}) <2/day$  and $p_{astro} > 0.5 $ as per the GW transient catalogs, GWTC-2.1 \cite{theligoscientificcollaboration2022gwtc21} and GWTC-3 \cite{gwtc-3}) have PE posteriors readily available \footnote{The posterior samples are found in the data releases \cite{ligo_scientific_collaboration_and_virgo_2022_6513631} and \cite{ligo_scientific_collaboration_and_virgo_2021_5546663}.}, the subthreshold candidates generally do not. This is in part because there isn't sufficient evidence to suggest that they are of astrophysical origin and therefore worthy of PE follow-up. 

Moreover, acquiring PE samples for all the subthreshold candidates is computationally taxing, which makes most of the existing lensing-identification methods mentioned above computationally expensive and time-consuming to be feasibly used. This is true even for the low-latency approaches. PO requires PE posteriors on the intrinsic parameters of both events in the lensed pair candidate to be analyzed \cite{haris2018}.  On the other hand, the ML-based method needs extensive training and testing involving subthreshold events, which has yet to be completed \cite{goyal2021}. To mitigate the increase in candidate pairs by including subthreshold events, we introduce another method that rapidly constructs interpretable, albeit approximate, statistics to rank the candidate lensing counterparts to the super-threshold events that are found by the targetted subthreshold searches. The method then provides a {\it preliminary} identification of super-sub lensed candidate pairs.

Our method is akin to PO, although the data products used can be generated rapidly without taxing computational resources. In particular, the GW-likelihood is approximated as a Gaussian using a Fisher analysis \cite{cuttler-flanagan}. This enables a rapid, though approximate, construction of posterior distributions on the chirp masses of the super/ subthreshold events. For each super-sub candidate pair, the chirp-mass posteriors of the events in the pair are quantitatively compared using the Bhattacharyya distance \cite{bhattacharyya1946}. Another coefficient is produced by cross-correlating the Bayestar skymaps \cite{singer2016}, generated in low latency, of each of the events in the pair. A third coefficient, exploiting the expected time delay distribution of detected strongly lensed events, is also constructed. The coefficient values are then sorted in descending order to identify any super-sub candidate pair with large values across all three coefficients.

All super-sub candidate pairs were found to be insignificant by one or more of these statistics, with the notable exception of one solitary outlier, GW191230\_180458–LGW200104\_180425, which we call simply GW191230-LGW200104 from now on. This pair lies in the top 5th percentile of all super-sub pairs across all three statistics. The event pair was re-analyzed by PO and more sophisticated joint-PE methods \cite{golum,hanabi} and found to be among the more significant candidate pairs, though not sufficiently significant to claim observation of lensing. We point the reader to \cite{o3btechdoc} for more detailed analyses of this event pair.

The rest of the paper is organized as follows. Section \ref{sec:methods} describes the construction of the approximate statistics. Section \ref{sec:results} describes the results, including the performance of the approximate statistic on synthetic lensed and unlensed events as compared to PO, as well as the output of the method applied to the super-sub pairs. Section \ref{sec:conclusion} summarizes the paper and discusses the scope for future improvements.

%
%
%
%

\section{Methods}\label{sec:methods}

Given a pair of CBC events, we wish to determine whether they have a common provenance (lensed), or whether they are unrelated (unlensed). In the geometrical optics limit, each strongly lensed copy of the GW strain gets an overall magnification, time delay and a (Morse) phase shift, although the phase evolution remains unaffected. In addition, the angular separation of the images ($\mathcal{O}(1'')$ or smaller) is orders of magnitude smaller than what can be resolved with GW detectors (which provide sky areas of $\mathcal{O}(10)$ sq. deg or larger). 

Hence,  apart from luminosity distance, time of arrival, and coalescence phase, the Bayesian inference of intrinsic and extrinsic parameters should yield posterior distributions that overlap well. The posterior overlap statistic exploits this fact to distinguish between the lensed and unlensed candidate pairs. However, its reliance on the availability of PE posteriors makes it difficult to employ to identify sub-threshold image counterparts to super-threshold GW events. This is because PE posteriors are computationally expensive and time-consuming to produce and are thus usually unavailable for sub-threshold events.

 On the other hand, the matched-filter-based search pipelines give point estimates of the intrinsic parameters and the GW network's signal-to-noise ratio (SNR). Furthermore, the chirp mass of a CBC is expected to be the best measured among all intrinsic parameters, certainly for those events whose in-band signal is dominated by the inspiral. We use this to construct a statistic, $\mathcal{B}^\mathrm{masses}$, that estimates a ``distance'' separating the chirp-mass posteriors evaluated using a Fisher analysis. 
 
 Additionally, we use the Bayestar \cite{singer2016} sky localization software to rapidly re-construct the localization skymaps. We then cross-correlate them to evaluate another statistic, $\mathcal{B}^\mathrm{sky}$, that measures the degree of overlap between these skymaps.

 And finally, we use the $\mathcal{R}_{\mathrm{gal}}$ \cite{haris2018, More:2021kpb} statistic to assess if the time delay of the candidate is more consistent with the distribution of temporal separations of randomly distributed events within the observation time, or the distribution of simulated image time-delays constructed from an assumed distribution of galaxy lenses and sources.

\subsection{Bayestar Skymaps}    \label{subsec:skymaps}

``Bayestar" is a sky localization software that can produce skymaps in seconds, by exploiting the fact that the intrinsic and extrinsic parameters of a CBC are semi-uncorrelated. It pins the intrinsic parameters to their matched-filter-search point estimates and rapidly marginalizes out the remaining nuisance parameters using Gaussian quadrature methods, to provide a posterior on the right ascension ($\alpha$) and declination ($\delta$) of the source.  \cite{singer2016}

Following \cite{goyal2021}, we project the Bayestar skymaps to a Cartesian grid of the sky coordinates ($\alpha, \delta$).  With this probability density $p(\alpha,\delta \mid d_i)$ for each of the two images (where $d_i$ is the data pertaining to each of the images),  we evaluate the following overlap integral, motivated by the posterior overlap statistic \cite{haris2018}:
\begin{equation}
\mathcal{B}^\mathrm{overlap} = \int \frac{p(\vec{\theta} \mid d_1)p(\vec{\theta} \mid d_2)}{p(\vec{\theta})}d\vec{\theta},
\label{eq:PO}
\end{equation}
with $\vec{\theta} = \{\alpha , \delta\}$, assuming isotropic sky priors:  $p(\alpha, \delta) \propto \alpha \sin \delta $.  This is our first statistic, which we call $\mathcal{B}^\mathrm{sky}$.\\

\subsection{Chirp Masses}    \label{subsec:masses}

The search pipelines report the detector frame matched-filtered chirp mass, $\mathcal{M}_c^\mathrm{det}$, and SNR  for the triggers based on the best match template at each detector. Following \cite{Caliskan:2022wbh} we construct a posterior on the chirp mass for each image,  $p(\mathcal{M}_c^\mathrm{det} \mid d_i)$ as a Gaussian with mean as the average of matched-filtered chirp masses over detectors and standard deviation as:
\begin{equation}
\Delta (\log \mathcal{M}_c^\mathrm{det} ) = 0.08 (\rho_\mathrm{thresh})/ \rho
\label{eq:stdmasses},
\end{equation}
where $\rho$ is the network SNR for the triggers and $\rho_\mathrm{thresh} = 8$.  The Bhattacharyya coefficient in chirp mass ($\mathcal{B}^\mathrm{masses}$) is defined as, 
\begin{equation}
\mathcal{B}^\mathrm{masses} = \int \sqrt{P(\mathcal{M}_c^\mathrm{det}|d_1)P(\mathcal{M}_c^\mathrm{det}|d_2)}d\mathcal{M}_c^\mathrm{det},
\label{eq:bmasses}
\end{equation}
which is our second statistic for rapid lensing identification. Note that unlike the PO statistic (Eq. ~\ref{eq:PO}), the Bhattacharyya coefficient is a prior-independent measure and is normalized ranging from 0 to 1.  The integral in Eq. ~\ref{eq:bmasses},  is simplified for the Gaussian posteriors and can be written in terms of means and standard deviations of the two Gaussians. 

In Gaussian noise, the matched-filter SNR is an optimal statistic. However, non-Gaussianities in real noise can produce spuriously large SNRs. Furthermore, even in Gaussian noise, noise fluctuations, and discreteness of the template bank, could result in signals being recovered by templates whose parameters are significantly biased with respect to the true source parameters.  

\begin{figure}[htb]
\includegraphics[width=\linewidth]{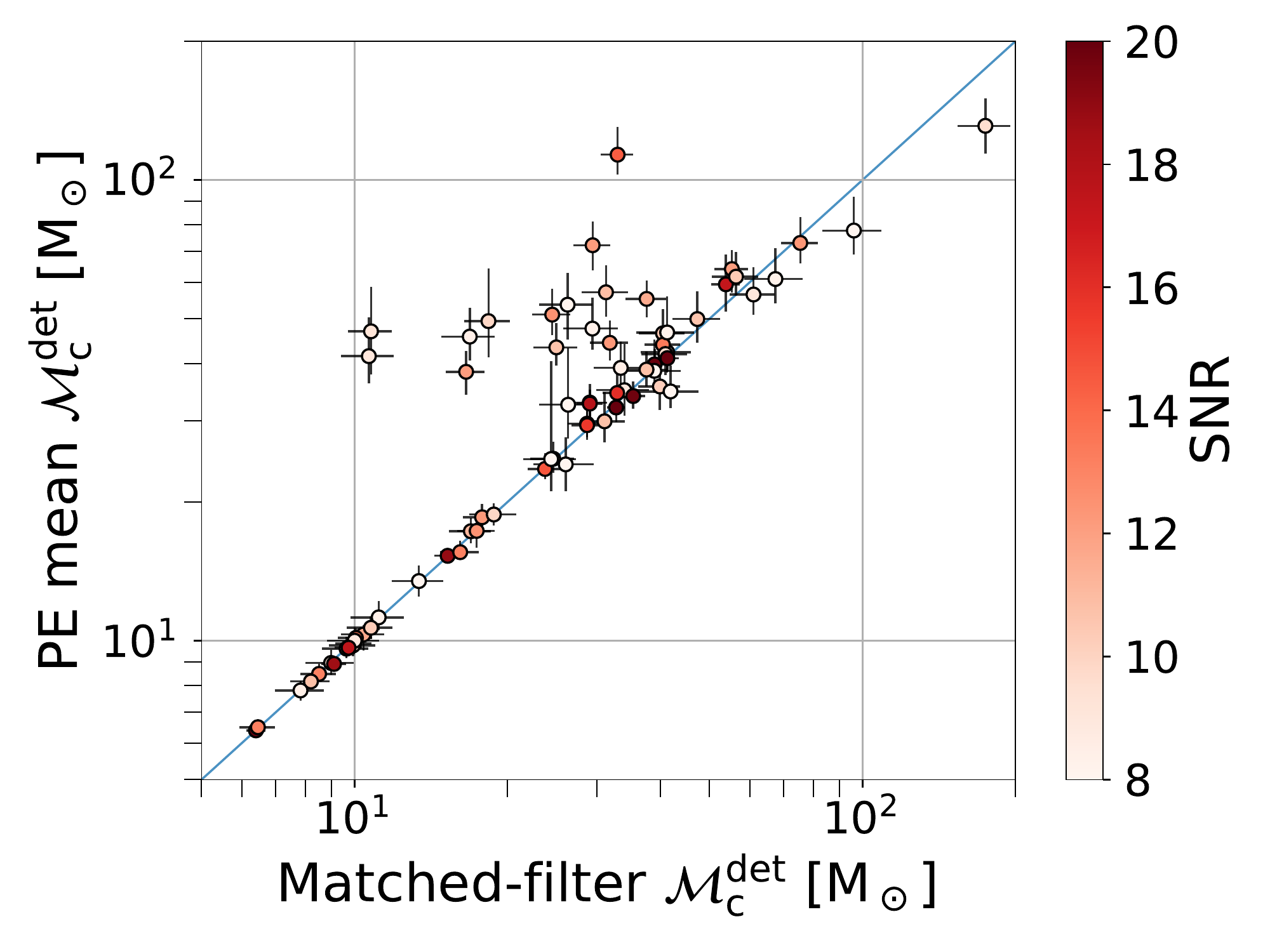}
\caption{Chirp mass estimates from PE and matched-filter-based searches for the GWTC-3 events. About $10-15 \%$ of events are significantly biased -- the mean value of the detector-frame chirp mass lying outside the $90\%$ confidence of the corresponding PE posteriors (vertical error bars) and matched-filter chirp mass (horizontal error bars) coming from Eq. \ref{eq:stdmasses}. If these error bars cross the diagonal, then those events should not be considered as significantly biased. The lower chirp mass binaries are less biased than the high ones due to their longer inspiral in the frequency band of LIGO-Virgo detectors.}
\label{fig:biases}
\end{figure}

We compare in Fig. \ref{fig:biases} the estimates of chirp mass from PE with the matched filter estimates, for GWTC-3 events. The PE estimates are taken from GWOSC \cite{theligoscientificcollaboration2023open} whereas the matched filter estimates are taken from GraceDB, as reported by the search pipelines \cite{pycbc_software, cannon2020gstlal, Usman:2015kfa}. About 13 out of 81 events have $\mathcal{M}_c^{det}$ from the  matched-filter values significantly different (mostly overestimated) with respect to the PE estimates. This is seen in the figure as points sufficiently deviated from the diagonal such that neither the vertical nor the horizontal error bar intersect the diagonal. Most of these events are in the chirp mass range $40-60 M_\odot$.  We do not find any correlation with the SNR for these biases. In addition to the chirp mass, the matched-filter searches also output the best-fit template's mass ratio and spins. This information can in principle be considered for lensing identification however, we leave it for future work.

The lower chirp mass binaries are less biased than the high ones due to their longer inspiral in the frequency band of LIGO-Virgo detectors. This is clearly a caveat but is partially mitigated by the fact that the Bhattacharyya coefficient is not used in isolation but in conjunction with other statistics in this section which are not susceptible to template biases.

\subsection{ Time-Delay Distributions}    \label{subsec:td}

The time of arrival of GW signals is measured at $\mathcal{O}$(ms) precision by matched-filter searches. It is therefore worthwhile to construct a statistic that uses arrival times as a means to discriminate between lensed and unlensed events. To that end, an $\mathcal{R}^\mathrm{gal}$ statistic is constructed.

The distribution of time delays for unlensed pairs of GW signals $P(\Delta t| H_U)$ can be estimated by assuming the arrival times to follow a Poisson process (see Eq. 31 in \cite{hanabi}).  Conversely, the distribution of time delays for detectable lensed pairs depends on the distribution of lens parameters, source parameters, and the relative separations between the Earth, the lenses and the sources. 

For galaxy-scale lenses, the time delays could vary from several minutes to several weeks. For cluster-scale lenses, the time delays could even span months and years. Assuming an appropriate model for the distribution of galaxy-lens parameters, as well as the redshift distributions of lenses and sources, we can construct a distribution of time delays $P(\Delta t| H_L)$ \cite{More:2021kpb} pertaining to detectable strongly lensed BBHs. 

In particular, we assume to be the singular isothermal ellipsoid (SIE). The parameters, including redshifts, of the lens, are assumed to follow the SDSS catalog \cite{Choi:2007a}. The BBH mergers are distributed over redshift following Oguri et. al. \cite{Oguri:2018muv}. The mass spectrum of BBHs is assumed to be the power-law + peak model \cite{gwtc3-population}. We set the detectability criteria to be network $SNR > 8$.\footnote{The sub-threshold events in O3 have $SNR > 7$ (see fig. \ref{fig:O3_snr}) however, the time delay distributions do not change noticeably with this choice.}

\begin{figure}
\includegraphics[width = \linewidth]{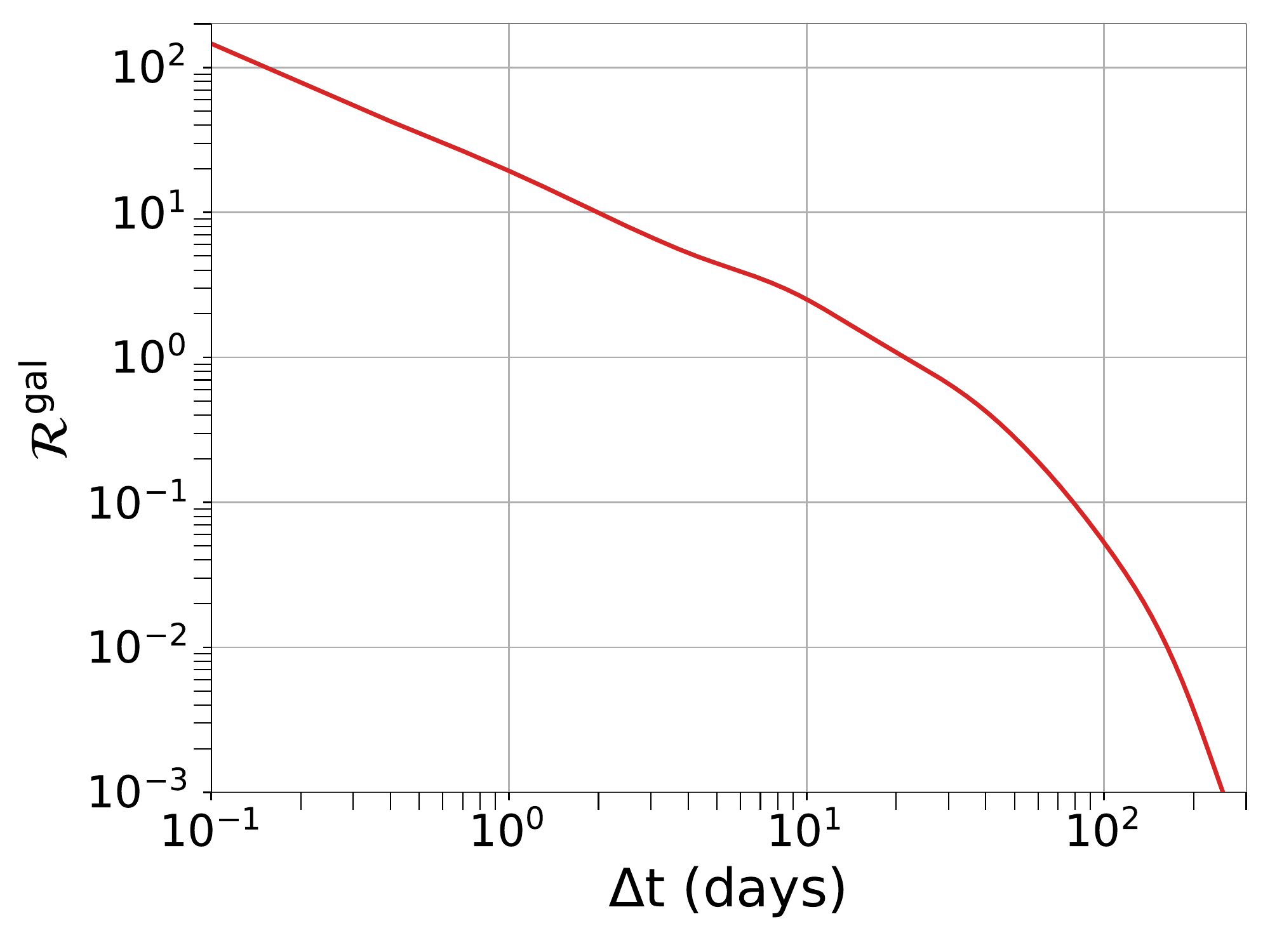}
\caption{$\mathcal{R}^\mathrm{gal}$ statistic as a function of the time delay between the events, assuming the O3 observation time. This is a model-dependent statistic -- the galaxy lens is assumed to be a singular isothermal ellipsoid, and the lens parameters are fit to the SDSS catalog. Lens redshifts are also assumed to follow the SDSS catalog, while source masses assume a power-law + peak model.}
\label{fig:rgal}
\end{figure}

The $\mathcal{R}^\mathrm{gal}$  statistic is constructed from time-delay distributions of lensed and unlensed events as \cite{haris2018}:

\begin{equation}
\mathcal{R}^\mathrm{gal} = \frac{P(\Delta t_0| H_L)}{P(\Delta t_0| H_U)},
\end{equation}
where $\Delta t_0$ is the measured time delay between a given pair of signals.  Fig.  \ref{fig:rgal} shows $\mathcal{R}^\mathrm{gal}$ statistic as a function of the time delay between the events, assuming the observation time of the full O3 run.  The $\mathcal{R}^\mathrm{gal}$ statistic favors small time delays and falls off rapidly with increasing time delays. Though this is a model-dependent statistic and valid only for galaxy lenses, it improves our capability of identifying lensed events from unlensed ones \cite{haris2018}.

\section{Results}\label{sec:results}
\subsection{Assessment of the method with simulations}
To assess the performance of our method, we first apply it to simulated lensed and unlensed events. These are injected in Gaussian noise, generated using the zero-detuned high-power PSDs of Advanced LIGO and Advanced Virgo at their design sensitivities \cite{AvirgoSensitivity,aLIGOSensitivity}, as implemented in $\textsc{pycbc}$ \cite{pycbc, pycbc_software}. We compare the performance of our method with the PO statistic, using the injection set given in ~\cite{haris2018} which consists of roughly 300 lensed pairs and half a million unlensed pairs.

To calculate $B^\mathrm{masses}$, we assume the posteriors in $\mathcal{M}_c^\mathrm{det}$ to be Gaussian with mean as the maximum likelihood estimates from the PE runs (already performed over the mentioned injection set for the results of ~\cite{haris2018})  and use the SNR of the injection to calculate the standard deviation (see Eq. ~\ref{eq:stdmasses}). We compare the performance of $B^\mathrm{masses}$ to the PO statistic $\mathcal{B}^\mathrm{overlap}$ evaluated using the posteriors in component mass,  setting $\vec \theta = \{m_1,m_2\}$ in Eq. \ref{eq:PO}. The maximum likelihood estimates of the masses are a proxy for the matched-filter search estimates -- a proxy that is expected to be a good approximation for Gaussian noise.
For a templated search involving real noise,  apart from the non-Gaussian nature of the latter, we would have an additional source of error in chirp mass estimation incurred due to template-bank discreteness.

To construct $B^\mathrm{sky}$, we generate the Bayestar skymaps using the same injection parameters and noise properties as used in the PE runs mentioned above, and fixing the intrinsic parameters to their true values. We compare $B^\mathrm{sky}$ to the PO statistic $\mathcal{B}^\mathrm{overlap}$ evaluated using the PE posteriors in sky location,  setting $\vec \theta = \{\alpha,\delta\}$ in Eq. \ref{eq:PO}. 

\begin{figure}[htb]
\includegraphics[width=\linewidth]{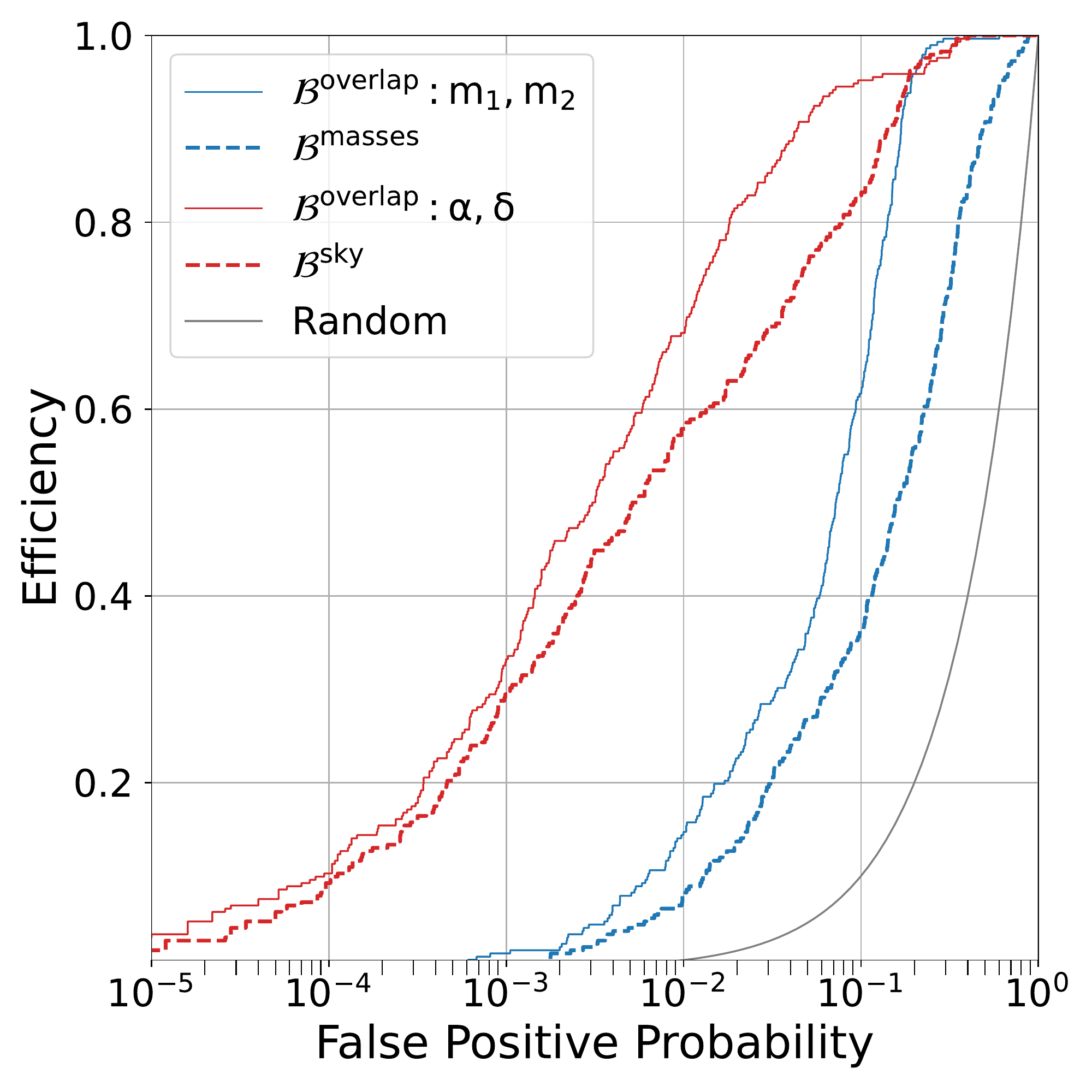}
\caption{Comparison of the rapid identification statistics against the PE-based posterior overlap method. There is only a partial loss in efficiency using these statistics relative to PO. }
\label{fig:harisetal}
\end{figure}

We plot the receiver-operating-characteristics (ROCs) graphs in Fig. \ref{fig:harisetal} for each statistic. The ROCs display the efficiency (i.e. the fraction of lensed events truly identified as lensed) at a given false positive probability (FPP, the fraction of unlensed events falsely identified as lensed).  

An ideal classifier would have an efficiency of 1 for all FPPs. However, during lensing identification, false positives can arise due to chance overlaps of the posteriors of unrelated events.  Hence as we increase the threshold of a statistic to identify a pair as lensed, the number of false positives reduces but at the cost of reducing the efficiency. As seen in the figure, both the mass-based and skymap-based statistics produce reduced efficiencies relative to the posterior overlap method. 


\subsection{O3 targeted sub-threshold search triggers}\label{subsec: TESLA}
In the case of strongly-lensed gravitational waves, a sub-threshold event with the same intrinsic parameters as the super-threshold event is expected. We consider the \texttt{GstLAL}-based TargetEd Sub-threshold Lensing seArch (TESLA) method \cite{Li:2019osa} to search for sub-threshold lensing counterparts. This is done using the posterior samples from targeted super-threshold event to reduce the background noise. Since the lensed counterpart can be de-magnified and obscured by noise, making it difficult to detect, it is necessary to minimize the impact of background noise. By reducing the amplitude and thus the optimal SNR, we generate these signals and inject them into actual data, which is analyzed using TESLA with a general template bank used in O3. We retain the templates that detect these injected signals and construct a reduced targeted bank. Finally, this targeted bank is used to identify potential lensed candidates for the targeted event from all possible data. For interested readers, please find the detailed description in \cite{Li:2019osa}.
 
As part of the LVK-collaboration wide paper to search for lensing signatures in the full third observing \cite{theligoscientificcollaboration2023search}, TESLA method was applied to all super-threshold events with a probability of astrophysical origin $p_\mathrm{astro}>0.5$ documented in \cite{gwtc-3} to search for their possible sub-threshold lensed counterparts, should they exist. For each targeted search, we keep all candidates with a $\mathrm{FAR} < 1$ in $30$ days (i.e. $<3.86\times10^{-7}$Hz) and that pass a preliminary skymap overlap test\cite{Wong:2021lxf} (See \cite{theligoscientificcollaboration2023search} for details). 

\begin{figure}[htb]
\includegraphics[width = \linewidth]{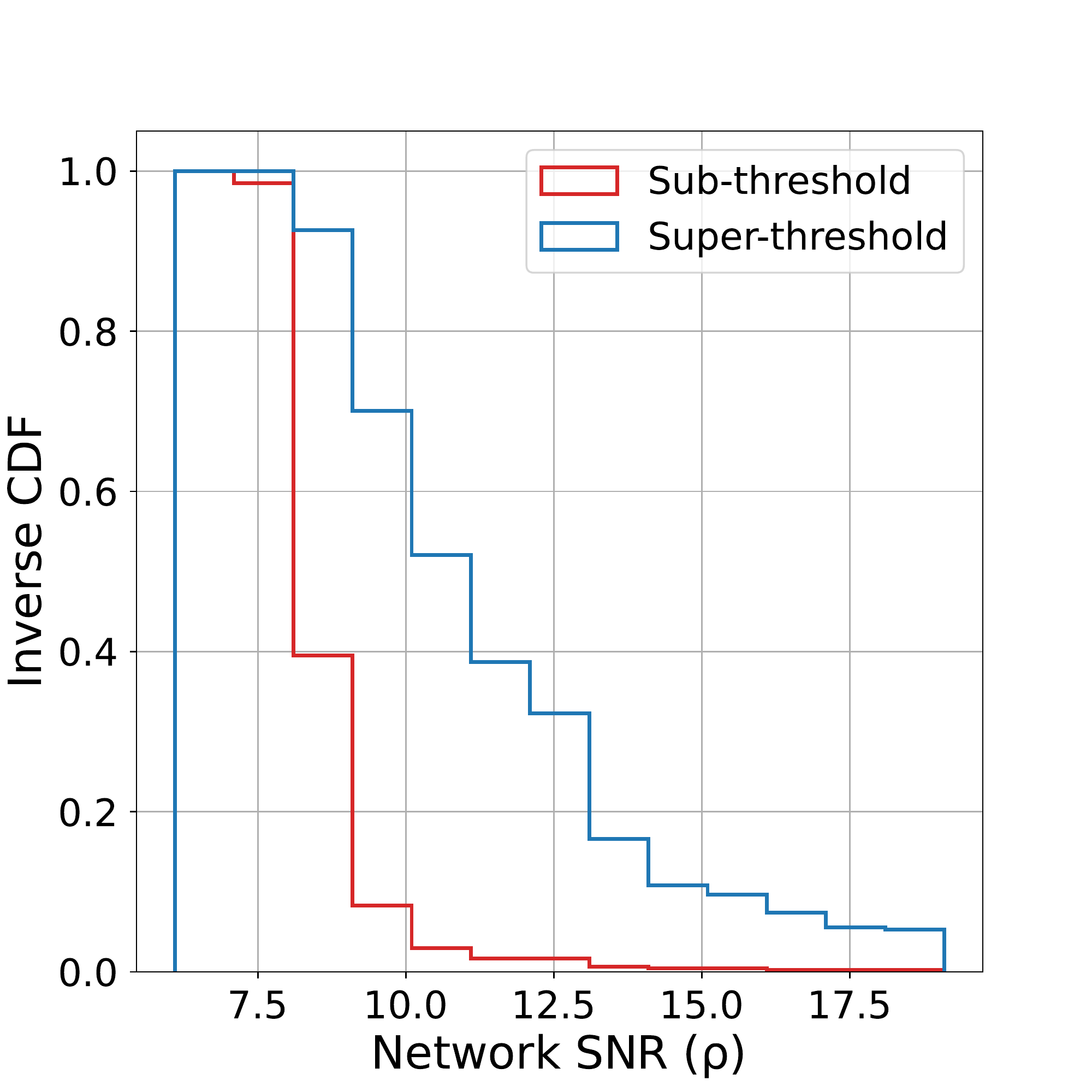}\\

\caption{Inverse cumulative distribution function  of the network SNRs. The sub-threshold triggers have lower SNRs as compared to super-threshold ones but all of them have $ \mathrm{SNR} > 7$. In fact, $40\%$ of the subthreshold events have $\mathrm{SNR} > 8$.}
\label{fig:O3_snr}
\end{figure}

For all the O3 events, altogether $472$ possible sub-threshold lensed candidates were found as a deeper internal candidate list \footnote{The candidates having $90\%$  credible region skymap overlap $> 0$, taken from table 1 of the data released in \cite{ligo_scientific_collaboration_and_virgo_2023_7863629}} of which only a small subset is reported in table I \cite{theligoscientificcollaboration2023search}. Fig.  ~\ref{fig:O3_snr} shows the distribution of network SNRs of those superthreshold and subthreshold events.  As expected the subthreshold signals have lower SNRs as compared to superthreshold ones. Nevertheless, they all have $\mathrm{SNR} > 7$.  Moreover, ~$40\%$ of the subthreshold events have $\mathrm{SNR} > 8$. This is not necessarily surprising, given that in real noise, SNR is known to be suboptimal, relative to its performance in Gaussian noise. A more robust statistic, $p_{astro}$ \cite{Farr:2013yna, Kapadia:2019uut}, is therefore used to segregate signals of astrophysical and terrestrial origin.

\subsection{Rapid Identification of the super-sub lensed candidates in O3}

During the LVK full O3 lensing searches \cite{theligoscientificcollaboration2023search},  only 2 out of all the above-mentioned targeted super-sub lensed pair candidates, pertaining to subthreshold events with the lowest $\mathrm{FAR}$ (i.e rate at which noise can falsely trigger a GW-like event) were followed up by a joint-parameter estimation analysis as implemented in the GOLUM pipeline \cite{golum}. None of those candidates showed any signatures of lensing. Here we consider all the $472$ search triggers and perform a preliminary analysis to rapidly identify the most interesting super-sub lensed candidates using the matched-filter estimates and the Bayestar skymaps. The chirp mass and skymaps overlap for the GW event pairs is captured in the $\mathcal{B}^\mathrm{masses}$ and $\mathcal{B}^\mathrm{sky}$ respectively (see sec. ~\ref{sec:methods}).  

\begin{figure}[htb]
\includegraphics[width =  \linewidth]{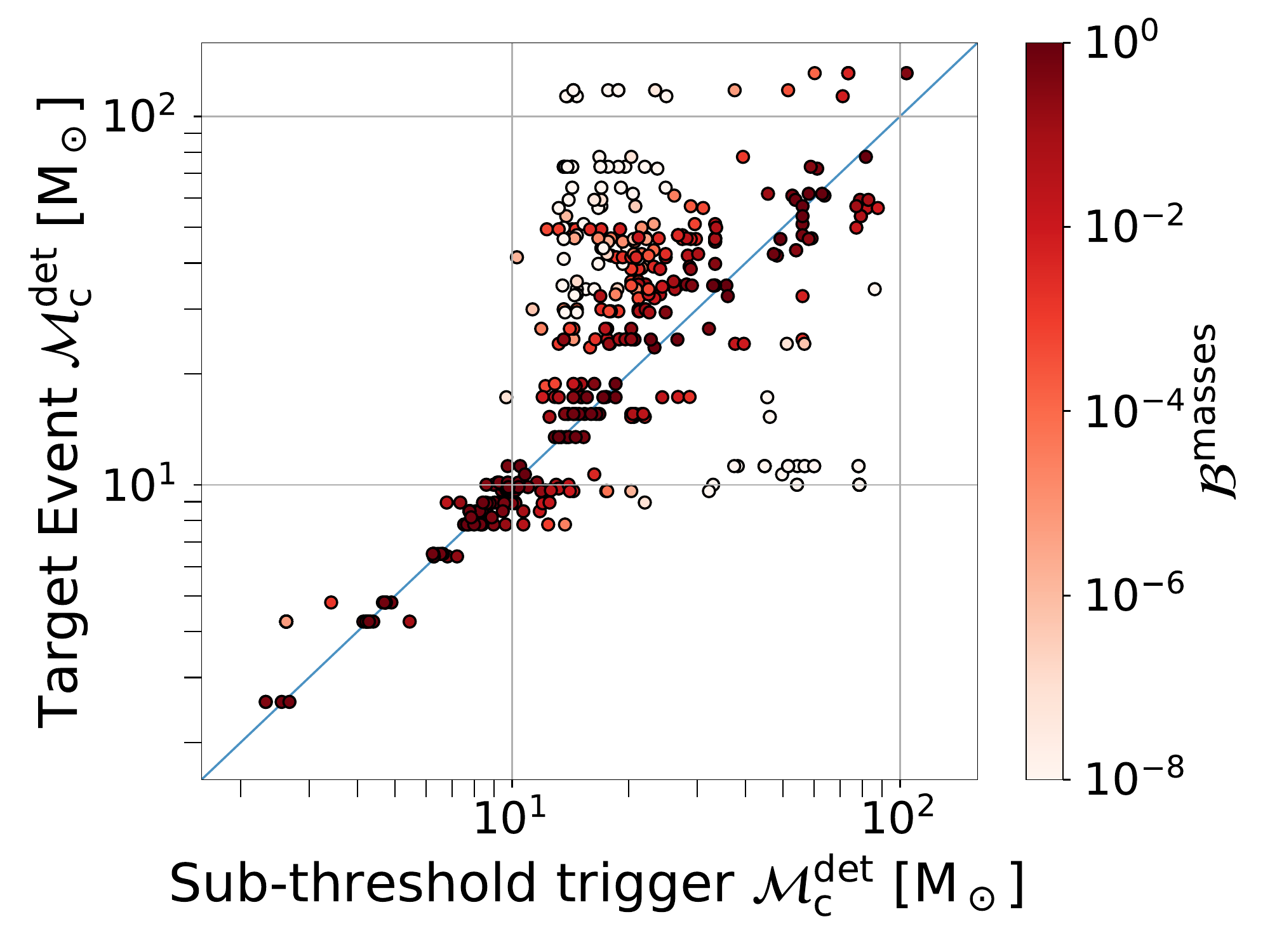}
\caption{Mean chirp mass estimates for the O3 super-sub candidate pairs and their $\mathcal{B}^\mathrm{masses}$. As expected, the majority of events with large $\mathcal{B}^\mathrm{masses}$ lie along the diagonal. The coefficient's value decreases for events situated away from the diagonal.}
\label{fig:O3_bc}
\end{figure}

In order to calculate the chirp mass overlap statistic $\mathcal{B}^\mathrm{masses}$, for the O3 super-sub pairs we use the matched-filter estimates of  $\mathcal{M}_{c}$  and $\rho$ for the sub-threshold triggers,  whereas for the super-threshold events we use the PE-based measurements from the open data available in GWOSC \cite{RICHABBOTT2021100658, theligoscientificcollaboration2023open}. Note that PE and matched-filter chirp mass estimates can deviate significantly for a small fraction of events, as shown in Fig.~\ref{fig:biases}.

Assuming a Gaussian distribution of the chirp mass posteriors, using Eqs. \ref{eq:stdmasses}-\ref{eq:bmasses}, we calculate the Bhattacharyya coefficient $\mathcal{B}^\mathrm{masses}$ for each of the super-sub pairs.  Fig.~\ref{fig:O3_bc} shows the mean estimates of chirp masses for the O3 super-sub pairs and their corresponding $\mathcal{B}^\mathrm{masses}$.  As expected, the $\mathcal{B}^\mathrm{masses}$ is maximum along the diagonal i.e.  when the masses of the super-sub pair are similar.  Note that $\mathcal{B}^\mathrm{masses}$ depends both on the mean and  on the standard deviation of the two Gaussians, therefore as we move away from the diagonal it falls off but not monotonically.

\begin{figure}
\includegraphics[width = \linewidth]{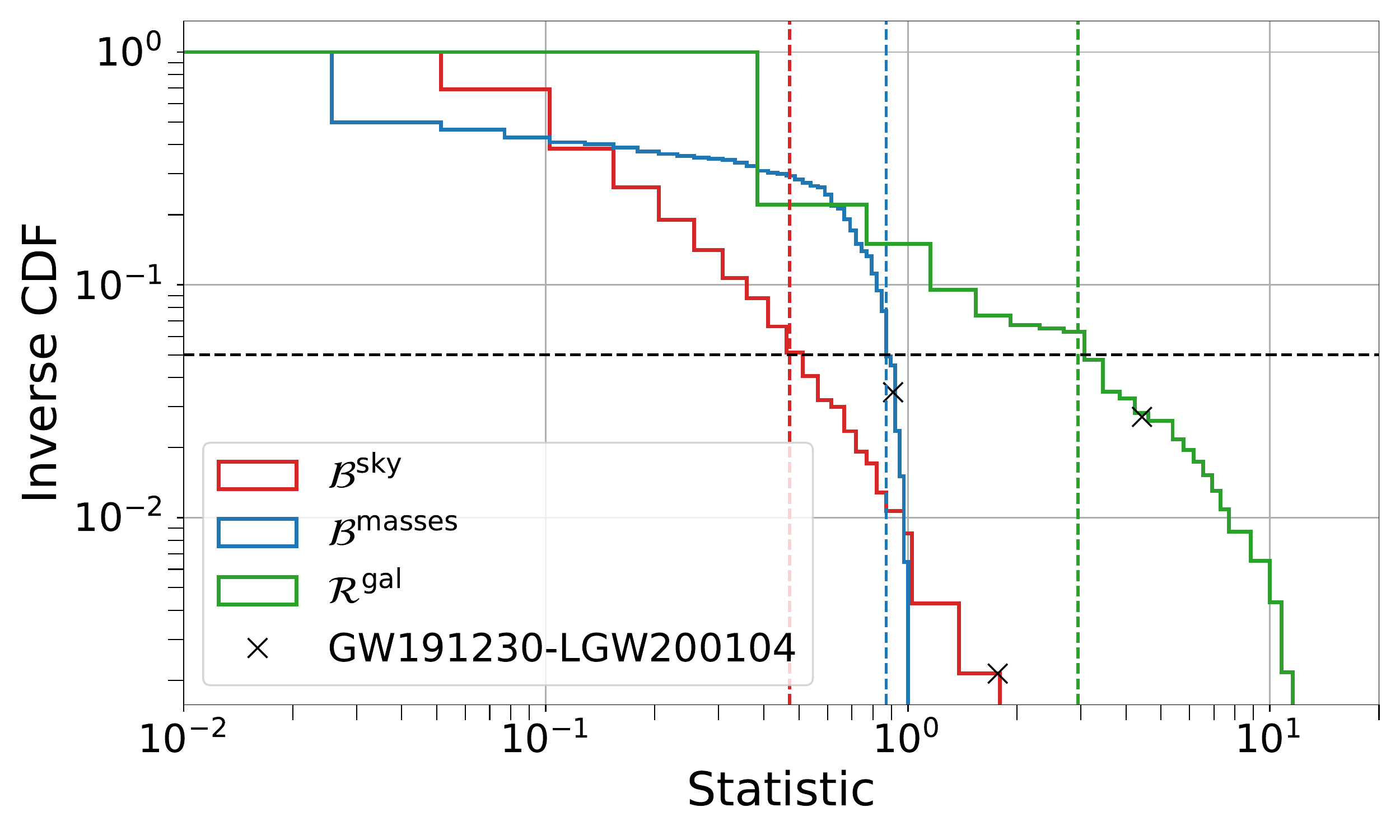}	
\caption{Inverse cumulative distribution function of the statistics for the O3 super-sub pairs. The dashed lines correspond to the top 5 percentile of the estimated values, which is found to be  at $\mathcal{R}^\mathrm{gal} = 2.95$, $\mathcal{B}^\mathrm{sky} = 0.47 $, and $\mathcal{B}^\mathrm{masses} = 0.87$ respectively. Only one event lies within this percentile, across all three statistics which is shown as a cross-mark ($\times$).}
\label{fig:stat_dist}
\end{figure}

\begin{figure*}[htb]
\includegraphics[width = \linewidth]{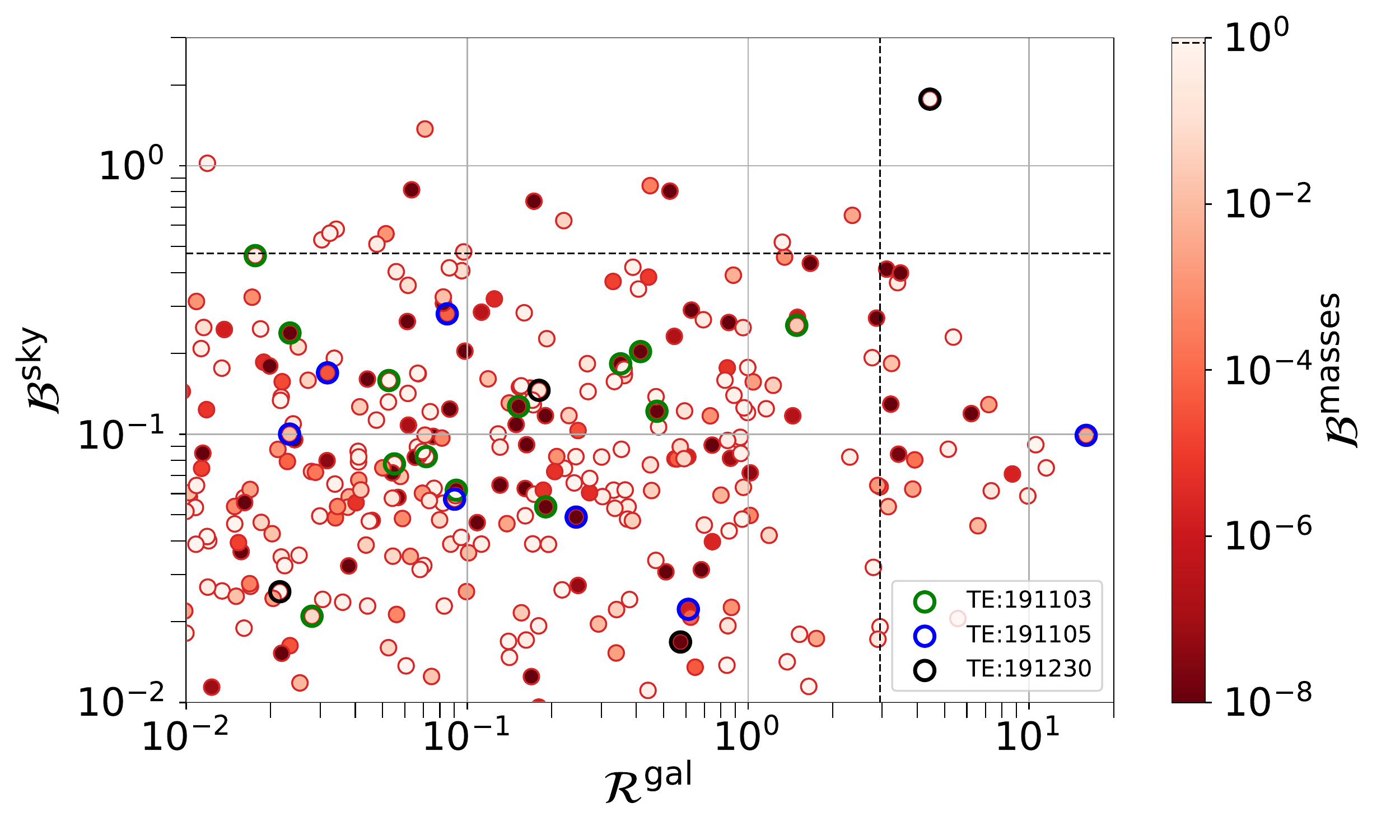}	
\caption{Statistics for the individual super-sub pairs which were found by TESLA during the LVK full O3 lensing searches \cite{theligoscientificcollaboration2023search}.  Black dashed lines represent the statistic value corresponding to the top 5 percentile of all the pairs. An obvious, solitary, outlier is identified in the top-right corner of the plot. In addition to being in the top 5th percentile of the two statistics on the x and y axes, it is also in the top $5^\mathrm{th}$ percentile of the statistic on the color bar (see the dashed black line there). We also highlight the triggers corresponding to the target events GW191230 (green), GW191105 (blue) and GW191103 (black).}
\label{fig:O3}
\end{figure*}


Next, we calculate the sky overlap statistic $\mathcal{B}^\mathrm{sky}$ (Eq. \ref{eq:PO}), using the Bayestar skymaps of the sub-threshold events and the more accurate PE skymaps of the super-threshold events.  Finally, we estimate the $\mathcal{R}^\mathrm{gal}$ by using the trigger time information of the events.  Fig. \ref{fig:O3} shows the time delay statitic $\mathcal{R}^\mathrm{gal}$, chirp-mass overlap $\mathcal{B}^\mathrm{masses}$ and sky-overlap $\mathcal{B}^\mathrm{sky}$ statistics. for each of the super-sub pairs. The dashed lines correspond to the top 5 percentile statistic values of all the O3 pairs, which are found to be at $\mathcal{R}^\mathrm{gal} = 2.95$, $\mathcal{B}^\mathrm{sky} = 0.47 $, and $\mathcal{B}^\mathrm{masses} = 0.87$ respectively. These are estimated from the distribution of the statistics for O3 super-sub pairs as shown in Fig. \ref{fig:stat_dist}.  Since lensing is a rare event (rates of strong lensing vary from $0.01\%$ to $0.1\%$ \cite{Xu2021, Wierda_2021}),  we want to find the event pairs which are in the tails of the distribution of the three statistics as the lensing candidates. It should however be noted that the analysis could suffer from spurious biases (see Fig. \ref{fig:biases}) and hence should be followed up by a more comprehensive analysis.

\begin{figure}[tbh]
    \centering
    \includegraphics[width=\linewidth]{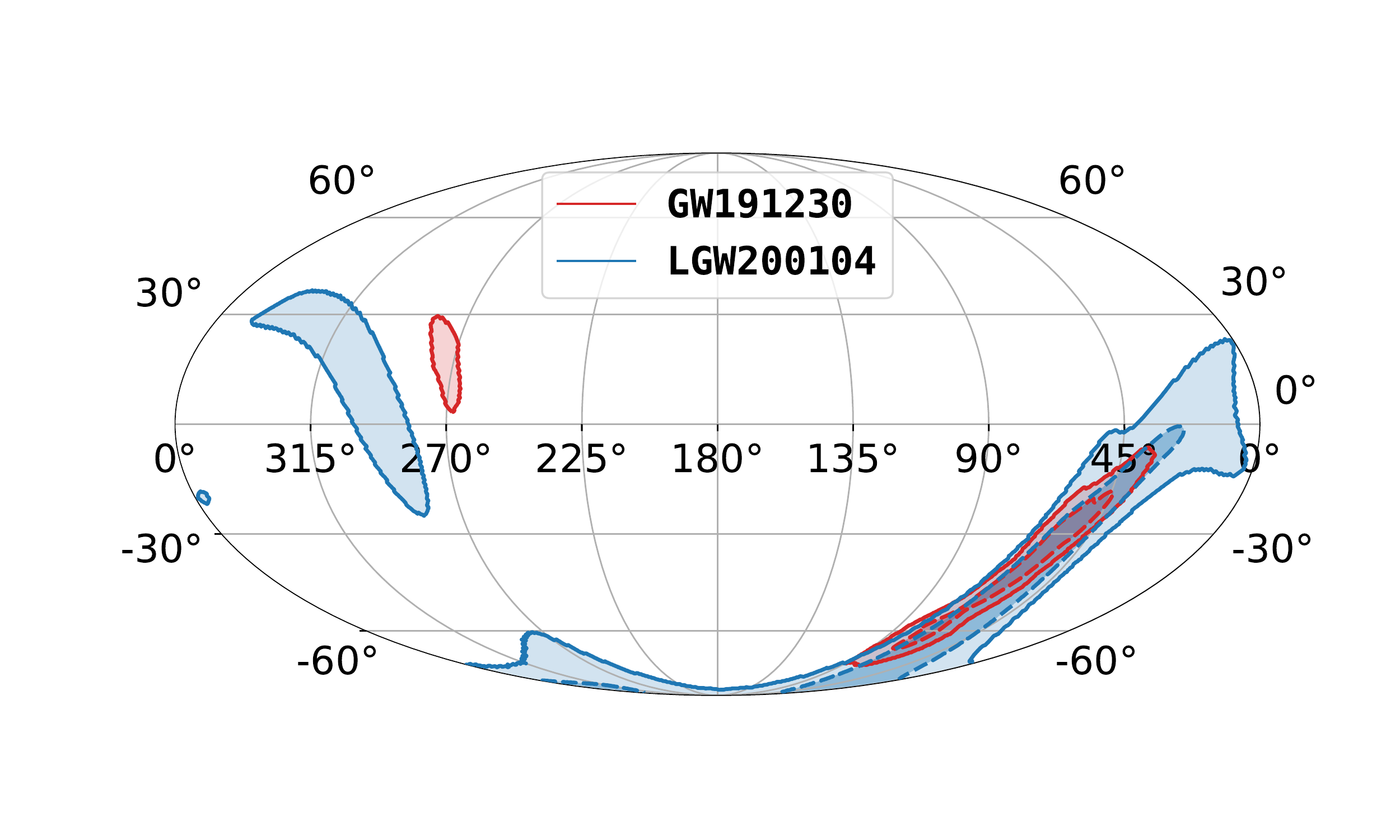}
    \caption{Skymaps of GW191230 and LGW200104. The dark (light) shaded region represents the 50 (90) \% contour. There is a significant visual overlap between the two skymaps. This is further corroborated by the high skymap overlap statistic value.}
    \label{fig:191230_sky}
\end{figure}

The only pair which is in the top 5 percentile of all three statistics is GW191230-LGW200104,  with LGW200104 as the sub-threshold counterpart to the super-threshold target event GW191230.  Moreover, this pair has a time delay of only $\sim 5$ days, which is more consistent with a lensed time delay than the temporal separation of two unrelated events drawn randomly from a Poisson process.  The statistics values for this pair is $\mathcal{R}^\mathrm{gal} = 4.43$, $\mathcal{B}^\mathrm{sky} = 1.77$, and $\mathcal{B}^\mathrm{masses} = 0.9$. The good overlap of the skymaps for the pair is evident from Fig. \ref{fig:191230_sky}. We conclude from our analysis that this pair is a (possibly lensed) outlier among the remaining population of (unlensed) O3 super-sub threshold pairs.

This pair was followed up by PE-based lensing analysis such as posterior overlap and joint-PE, in the \cite{o3btechdoc}. It turns out that after incorporating the lensing models and selection effects \cite{hanabi}, this pair has the highest significance (i.e.  highest probability of being lensed), amongst all the O3 pairs including the super-super threshold ones with a caveat that LGW200104 might be of terrestrial origin given that during the LVK online unlensed super-threshold searches, the event was found with the \textsc{SPIIR}~\citep{Luan:2011qx, Chu:2020pjv} and \textsc{cWB}~\citep{Klimenko:2015ypf} pipelines, which reported a $p_{astro}$ of 1\% and a $\mathrm{FAR}$ of $4824.6 /yr$. This is in contrast to the FAR of $6.59/yr$ estimated by TESLA. We report these numbers for the benefit of the reader, and do not attempt to draw any conclusions about the nature (noise or signal) of LGW200104. Our method gives a {\it preliminary} assessment of whether LGW200104 is a lensed counterpart to GW191230, {\it assuming} the former is a signal, which it well may {\it not} be. We follow the LVK naming convention of the GW candidates i.e.  $GWYYMMDD\_hhmmss$, encoding the date and Coordinated Universal Time (UTC) of the signal \cite{gwtc-3}. Table \ref{table:TE191230} shows the rest of the triggers found in the searches with the target event GW191230. All of them have at least one of the statistics whose value is low enough to be relegated as unworthy of follow-up.

\begin{figure}[tbh]
    \centering
    \includegraphics[width=\linewidth]{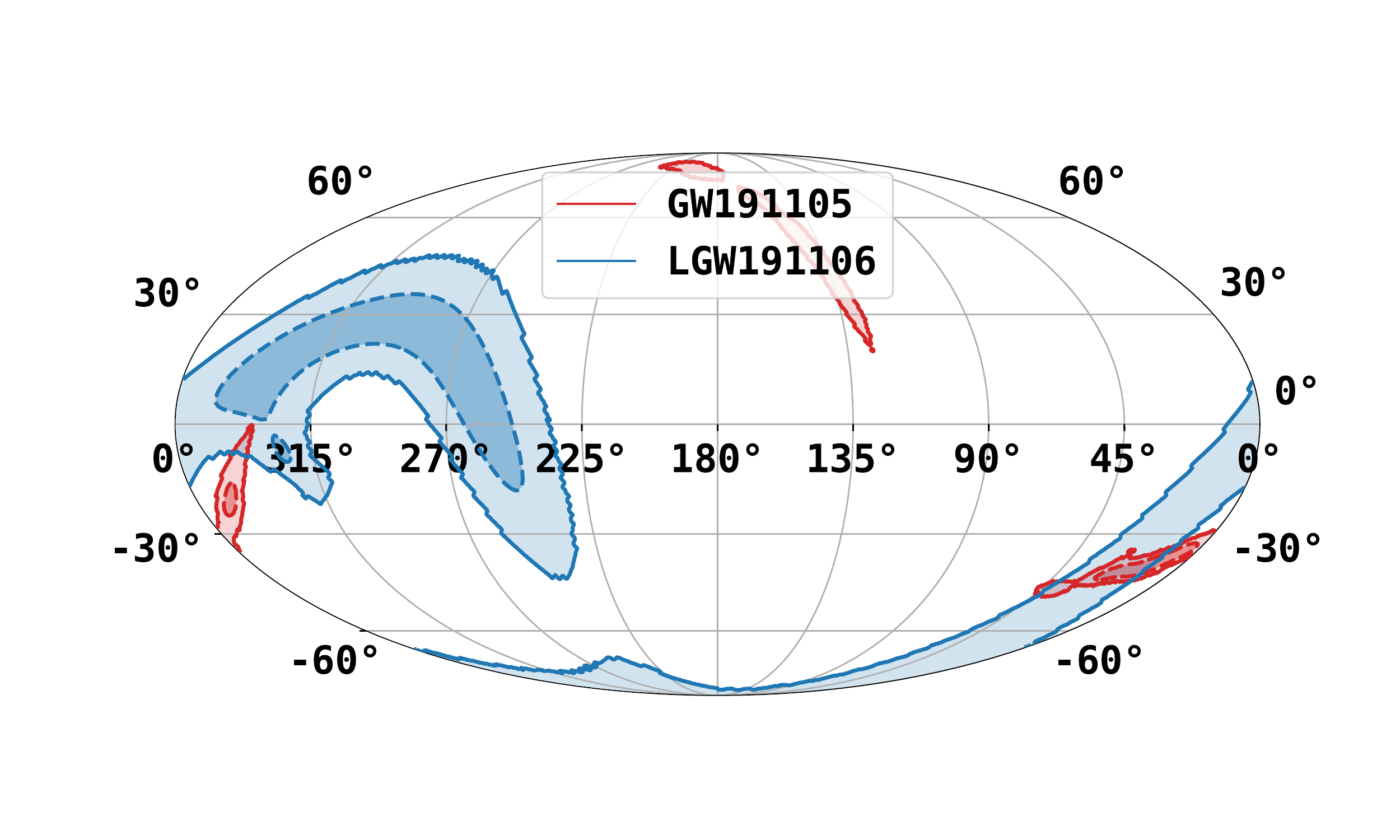}
    \caption{Skymaps of GW191105 and LGW191106. The dark (light) shaded region represents the 50 (90) \% contour. The skymap overlap is even visually seen to be modest and is also evidenced in the low skymap overlap statistic value.}
    \label{fig:191105_sky}
\end{figure}

\begin{table*}[htb]
\centering
\begin{tabular}{ c| c| c| c| c|c|c|c} 
\hline
Sub-threshold  Trigger & $\mathcal{M}_c^\mathrm{det}$ [$M_{\odot}$] & SNR & FAR $[\mathrm{yr}^{-1}]$ & $\Delta t$ [days] & $\mathcal{R}^\mathrm{gal} $& $\mathcal{B}^\mathrm{sky}$ & $\mathcal{B}^\mathrm{masses}$  \\

\hline
\textbf{LGW200104\_184028} &   58.20 & 8.48 &          6.59 &     5.02 & 4.43 &         1.77 &                       0.91 \\
LGW200301\_075426 &    45.70 & 7.15 &          5.61 &    61.58 & 0.18 &         0.15 &                       0.12 \\
LGW200201\_192756 &    20.50 & 7.80 &          5.11 &    33.06 & 0.57 &         0.02 &                       0.00 \\
LGW190818\_232544 &    63.00 & 8.51 &          3.34 &  -133.78 & 0.02 &         0.03 &                       0.99 \\
 \hline
\end{tabular}
\caption{Triggers for the target event GW191230\_180458, whose median  $\mathcal{M}_c^\mathrm{det}$ [$M_{\odot}$] = 61.68. \label{table:TE191230} The last three columns are the statistics calculated using the method proposed here (see Sec. \ref{sec:methods}). LGW200104\_184028 favours lensing through all three statistics. Rest of the triggers show no sufficient evidence  of being lensed counterparts. }
\end{table*}
\begin{table*}[htb]
\centering
\begin{tabular}{ c| c| c| c| c|c|c|c} 
\hline
Sub-threshold  Trigger & $\mathcal{M}_c^{det}$ [$M_{\odot}$] & SNR & FAR $[\mathrm{yr}^{-1}]$ & $\Delta t$ [days] & $\mathcal{R}^\mathrm{gal} $& $\mathcal{B}^\mathrm{sky}$ & $\mathcal{B}^\mathrm{masses}$  \\
\hline
\textbf{LGW191106\_200820} &    14.40 & 8.36 &          0.53 &     1.23 & 15.95 &         0.10 &                       0.00 \\
LGW200128\_115458 &    17.60 & 7.80 &          7.57 &    83.89 &  0.08 &         0.28 &                       0.00 \\
LGW191207\_050023 &    20.30 & 7.87 &          3.04 &    31.60 &  0.61 &         0.02 &                       0.00 \\
LGW191229\_024823 &    32.20 & 9.08 &         12.05 &    53.51 &  0.24 &         0.05 &                       0.00 \\
LGW200303\_074125 &    17.50 & 8.13 &          5.87 &   118.71 &  0.03 &         0.17 &                       0.00 \\
LGW200126\_135203 &    11.90 & 7.72 &          6.75 &    81.97 &  0.09 &         0.06 &                       0.17 \\
LGW200315\_070710 &    14.10 & 7.70 &         10.91 &   130.69 &  0.02 &         0.10 &                       0.01 \\
 \hline
\end{tabular}
\caption{Triggers for the target event GW191105\_143521, whose median $\mathcal{M}_c^{det}$ [$M_{\odot}$] = 9.62. \label{table:TE191105} The last three columns are the statistics calculated using the method proposed (see Sec. \ref{sec:methods}).  LGW191106\_200820 has significant the time delay statistic ($\mathcal{R}^\mathrm{gal}$), but is ruled out by the remaining two. Rest of the triggers show no signatures of being lensed counterparts. }
\end{table*}

\begin{table*}[htb]
\centering
\begin{tabular}{ c| c| c| c| c|c|c|c} 
\hline
Sub-threshold  Trigger & $\mathcal{M}_c^{det}$ [$M_{\odot}$] & SNR & FAR $[\mathrm{yr}^{-1}]$ & $\Delta t$ [days] & $\mathcal{R}^\mathrm{gal} $& $\mathcal{B}^\mathrm{sky}$ & $\mathcal{B}^\mathrm{masses}$  \\
\hline
LGW191118\_113217 &    14.00 &  8.08 &          6.02 &    15.42 & 1.49 &         0.25 &                       0.02 \\
LGW191213\_164018 &    78.60 & 10.40 &          0.32 &    40.64 & 0.41 &         0.20 &                       0.00 \\
LGW190919\_131654 &    33.00 &  9.15 &          2.89 &   -44.51 & 0.35 &         0.18 &                       0.00 \\
LGW190926\_133040 &    78.60 &  7.10 &          7.57 &   -37.50 & 0.47 &         0.12 &                       0.00 \\
LGW190828\_192315 &    54.30 &  7.37 &          6.24 &   -66.25 & 0.15 &         0.13 &                       0.00 \\
LGW200102\_033257 &    78.60 &  7.18 &          0.57 &    60.09 & 0.19 &         0.05 &                       0.00 \\
LGW200211\_024259 &     9.44 &  8.11 &          5.87 &   100.05 & 0.05 &         0.16 &                       0.81 \\
LGW200323\_135352 &     8.59 &  8.25 &          9.02 &   141.52 & 0.02 &         0.46 &                       0.33 \\
LGW190805\_134348 &    10.40 &  8.79 &          0.29 &   -89.49 & 0.07 &         0.08 &                       0.87 \\
LGW190813\_125024 &    78.60 &  7.14 &          5.61 &   -81.52 & 0.09 &         0.06 &                       0.00 \\
LGW200312\_144311 &    78.60 &  7.29 &          8.14 &   130.55 & 0.02 &         0.24 &                       0.00 \\
LGW190727\_144658 &     8.97 &  8.39 &          2.76 &   -98.44 & 0.06 &         0.08 &                       0.55 \\
LGW190517\_114359 &    78.60 &  7.40 &          0.85 &  -169.57 & 0.01 &         0.28 &                       0.00 \\
LGW190401\_190150 &    54.30 &  8.31 &          5.68 &  -215.27 & 0.00 &         0.36 &                       0.00 \\
LGW200305\_153119 &    13.00 &  8.61 &          4.92 &   123.59 & 0.03 &         0.02 &                       0.06 \\
\hline
\end{tabular}
\caption{Triggers for the target event GW191103\_012549, whose  median $\mathcal{M}_c^\mathrm{det}$ [$M_{\odot}$] = 10.01.\label{table:TE191103} The last three columns are the statistics calculated using the proposed method (see Sec. \ref{sec:methods}). None of the candidates shows signatures of being lensed.}
\end{table*}
{\it Special target events: GW191103\_012549 and GW191105\_143521}.  During the LVK full O3 strong lensing searches \cite{theligoscientificcollaboration2023search}, events pair GW191103\_012549--GW191105\_143521, which we call GW191103--GW191105 from now on,  was found to be among the more significant ($\sim 1\sigma$) pairs, though unlikely to be lensed. The pair is also analysed in detail in \cite{o3btechdoc} finding no conclusive evidence for lensing. Typically, galaxy lenses can produce more than two images of a GW source. Any additional sub-threshold lensed counterpart to the pair GW191103--GW191105 if found, can help us in increasing its significance of being lensed,  in the reconstruction of lens configuration and in localising the source to a host galaxy through cross-matching of electromagnetically observed lensed galaxy catalogues \cite{hannuksela2019, Wempe:2022zlk}. In the O3 data, 15 potential counterparts for GW191103 and 7 for GW191105 were found by the TESLA method (see Sec. \ref{subsec: TESLA}) but none of them is in common for both the targeted events.  The rapid statistics for each of them are shown in Tables \ref{table:TE191105} and \ref{table:TE191103}.   LGW191106\_200820 was also considered to be a potentially lensed counterpart of GW191105, by virtue of having the highest $ \mathcal{R}^\mathrm{gal}) = 15.95 $. However, the poor skymap overlap ($ \mathcal{B}^\mathrm{sky} = 0.1$) and even poorer Bhattacharya coefficient value resulted in this event being relegated as unworthy of follow-up. The skymaps for this pair are shown in Fig. \ref{fig:191105_sky}. In summary, we conclude that {\it no} obvious strongly lensed sub-threshold counterpart to the GW191103-GW191105 pair is found with our preliminary analysis.

\section{Summary and Outlook}\label{sec:conclusion}
A large number of targeted subthreshold events opens the possibility of one or more of them being lensed (de-magnified) counterparts of confidently detected superthreshold events. Identifying such subthreshold lensed counterparts using conventional methods involving large-scale parameter estimation exercises overburdens computational resources. Thus, a preliminary method that rapidly weeds out ``obviously" unlensed super-sub candidate pairs is required. In this work, we have proposed one possible rapid and computationally inexpensive identification scheme. The method is conceptually similar to the posterior overlap statistic while relying on approximations to posteriors on chirp masses and sky location pertaining to the subthreshold counterpart. 

A Bhattacharyya coefficient, $\mathcal{B}^\mathrm{masses}$, is constructed from the approximations to the chirp mass posteriors. A skymap overlap statistic, $\mathcal{B}^\mathrm{sky}$, is constructed from Bayestar skymaps of subthreshold events and PE skymaps of superthreshold events. The method additionally uses prior information on expected lensed time delays, assuming a lens profile and a SDSS-catalog-fitted model for lens parameter distributions, to further enhance its discriminating abilities. Accordingly, an $\mathcal{R}^\mathrm{gal}$ statistic is constructed.

Using this method, the vast majority of subthreshold events were found to be unrelated to the superthreshold events that were targeted. However, one interesting super-sub candidate pair was found to be an outlier -- its $\mathcal{B}^\mathrm{masses,~sky}$ and $\mathcal{R}^\mathrm{gal}$ statistic values were {\it all} found to be within the highest 5th percentile of the super-sub candidate pair values analyzed. The follow-up analysis of the pair in \cite{o3btechdoc} by more sophisticated joint-PE analyses, that also include selection effects \cite{hanabi}, deemed this event to be among the more significant lensed candidates among all analyzed candidate pairs -- super-super and super-sub, assuming a singular isothermal ellipsoid lens profile. While the significance of the event  was still not sufficient to claim the detection of a lensed pair in part because LGW200104 had a very low $p_{astro}$, the identification of this event using our method motivates the need for rapid and computationally inexpensive analyses for future observing runs where the number of candidate pairs is expected to grow drastically.

In future work, we intend to further assess the performance of our method in more realistic simulated data sets. In particular, we plan to inject subthreshold lensed and unlensed events, in addition to superthreshold events, in real noise. We then plan to perform a matched filter search for these events to acquire matched-filter parameter values of chirp masses and arrival times. Using these, and the Bayestar skymaps, we plan to reconstruct the ROCs to assess the performance in comparison to the posterior overlap statistic, to check if the loss in efficiency relative to PO is still acceptable. 

We additionally plan to explore the possibility of using PE products from PE algorithms that can rapidly generate posterior samples, such as heterodyning methods \cite{islam2022factorized, Roulet_2022, Cornish_2021} and machine learning based \cite{Dax_2021,Gabbard_2021, goyal2021} methods, among others \cite{pathak2022rapid}. These products will likely be a non-trivial improvement over those that were used in the method presented in this work, although this conjecture needs to be tested, especially for subthreshold events. 


\section{Acknowledgements} \label{sec:ack}

We are thankful to P. Ajith, D. Chatterjee, H. Fong and A. Barsode along with the astrophysical relativity group at ICTS and LVK lensing group for the helpful discussions and suggestions. We are also thankful to D. Keitel, M. Wright and J. Janquart for their careful reading of this manuscript. Juno C. L. Chan acknowledges support from the Villum Investigator program supported by VILLUM FONDEN (Grant No. 37766) and the DNRF Chair, by the Danish Research Foundation. J.R. Cudell benefits from the support of the Fonds de la Recherche Scientifique-FNRS, Belgium, under
grant No. 4.4503.19. 
The authors are grateful for computational resources provided by the LIGO Laboratory and supported by National Science Foundation Grants PHY-0757058 and PHY-0823459.

This research has made use of data or software obtained from the Gravitational Wave Open Science Center (gwosc.org), a service of LIGO Laboratory, the LIGO Scientific Collaboration, the Virgo Collaboration, and KAGRA. LIGO Laboratory and Advanced LIGO are funded by the United States National Science Foundation (NSF) as well as the Science and Technology Facilities Council (STFC) of the United Kingdom, the Max-Planck-Society (MPS), and the State of Niedersachsen/Germany for support of the construction of Advanced LIGO and construction and operation of the GEO600 detector. Additional support for Advanced LIGO was provided by the Australian Research Council. Virgo is funded, through the European Gravitational Observatory (EGO), by the French Centre National de Recherche Scientifique (CNRS), the Italian Istituto Nazionale di Fisica Nucleare (INFN) and the Dutch Nikhef, with contributions by institutions from Belgium, Germany, Greece, Hungary, Ireland, Japan, Monaco, Poland, Portugal, Spain. KAGRA is supported by Ministry of Education, Culture, Sports, Science and Technology (MEXT), Japan Society for the Promotion of Science (JSPS) in Japan; National Research Foundation (NRF) and Ministry of Science and ICT (MSIT) in Korea; Academia Sinica (AS) and National Science and Technology Council (NSTC) in Taiwan. This material is based upon work supported by NSF's LIGO Laboratory which is a major facility fully funded by the National Science Foundation.

\bibliography{references}

\end{document}